\documentclass[twocolumn,showpacs,preprintnumbers,amsmath,amssymb,superscriptaddress,pre]{revtex4}

\usepackage{graphicx}
\usepackage{dcolumn}
\usepackage{bm}

\begin{document}

\title{Mean field approaches to the totally asymmetric exclusion process with quenched disorder and large particles}

\author{Leah B.\ Shaw}
 \email{lbs22@cornell.edu}
 \affiliation{Department of Physics, Cornell University, Ithaca, NY 14853-2501}
 \affiliation{School of Chemical and Biomolecular Engineering,
 Cornell University, Ithaca, NY 14853-5201}

\author{James P.\ Sethna}
 \affiliation{Department of Physics, Cornell University, Ithaca, NY 14853-2501}

\author{Kelvin H.\ Lee}
 \affiliation{School of Chemical and Biomolecular Engineering,
 Cornell University, Ithaca, NY 14853-5201}

\date{\today}

\begin{abstract}
The process of protein synthesis in biological systems resembles a one
dimensional driven lattice gas in which the particles (ribosomes) have spatial
extent, covering more than one lattice site.  Realistic, nonuniform gene
sequences lead to quenched disorder in the particle hopping rates.  We study
the totally asymmetric exclusion process with large particles and quenched
disorder via several mean field approaches and compare the mean field results
with Monte Carlo simulations.  Mean field equations obtained from the
literature are found to be reasonably effective in describing this system. A
numerical technique is developed for computing the particle current rapidly.
The mean field approach is extended to include two-point correlations between
adjacent sites.  The two-point results are found to match Monte Carlo
simulations more closely.
\end{abstract}

\pacs{82.39.-k, 05.10.-a}

\maketitle

\section*{Introduction}

The process of protein synthesis, called ``translation,'' can be modeled using
a driven lattice gas in one dimension \cite{MGP, MG, LC, SZL}. This type of
model is well understood when all particle hopping rates are uniform, but a
model for the real biological process requires nonuniform particle hopping
rates. Direct Monte Carlo simulation of such models is possible when only a few
genes are involved.  However, it is desirable to perform large-scale
simulations to fit translation models to experimental data collected for many
genes simultaneously (e.g., data from \cite{B&B}). For this purpose, Monte
Carlo approaches would be computationally too slow. Therefore, other analytical
or computational methods are needed.

In this paper, we address the issue of quenched disorder (site-dependent
hopping rates) in a driven lattice gas model for translation.  The paper is
organized as follows.  The model is first described and its connection to
biological protein synthesis explained.  Section \ref{sec:literature} contains
a brief summary of known results.  In Section \ref{sec:coarsegrained}, we use a
simple coarse-grained approach to obtain crude, approximate solutions.  Section
\ref{sec:MGP} treats our central topic: application of a mean field method
\cite{MGP, MG} to the problem of quenched disorder. Analytical and
computational results are presented. In Section \ref{sec:twopoint}, we extend
the mean field approach to include two-point correlations for better accuracy.
We close with a brief summary and discussion of how these methods may be
applied to problems of interest, such as fitting translation models to
experimental data.

\section{Description of model}
\label{sec:model}

We focus on translation in prokaryotes, particularly \textit{Escherichia coli},
because of its relative simplicity. The process involves the synthesis of
specific proteins based on the sequence of messenger RNA (mRNA) molecules. The
mechanism consists of ribosomes ``reading'' the codons of mRNA as the ribosomes
move along an mRNA chain, and the recruitment and assembly of amino acids
(appropriate to the codons being read) to form a protein. (See, e.g., \cite
{Stryer}, for more details.) This process is often described as three steps:
initiation, where ribosomes attach themselves, one at a time, at the ``start''
end of the mRNA; elongation, where the ribosomes move down the chain in a
series of steps; and termination, where they detach at the ``stop'' codon.
Since ribosomes cannot overlap, their dynamics is subject to the excluded
volume constraint. Apart from being impeded by another ribosome (steric
hinderance), a ribosome cannot move until the arrival of an appropriate
transfer RNA, carrying the appropriate amino acid (a combination known as
aminoacyl-tRNA, or aa-tRNA). Thus, the relative abundances of the approximately
60 types \cite{NU} of aa-tRNA have significant effects on the elongation rate.
Assuming reactant availabilities in a cell are in their steady state, with a
time-independent concentration of ribosomes and aa-tRNA, there would be an
approximately steady current of ribosomes moving along the mRNA, resulting in a
specific production rate of this particular protein. Our goal is the prediction
of the protein production rates for various mRNA's, as a function of the
concentration of ribosomes and aa-tRNA's.

The process of translation is well suited to modeling using a driven lattice
gas in one dimension. Particles enter at some rate on one end of a chain of
discrete lattice sites (initiation), then hop down the chain one site at a time
with another rate or set of rates (elongation), and finally exit the chain at
the other end (termination). The excluded volume constraint is implemented by
insuring the spacing between ribosomes is no less than 12 sites, the
approximate number of codons that a ribosome blocks \cite{Heinrich, Kang}.
Quenched disorder arises because there is non-uniformity in the hopping
(elongation) rates along the chain.  This effect occurs because at each codon,
a ribosome has to ``wait'' for the appropriate aa-tRNA before continuing, and
the various aa-tRNA's are present in nonequal abundances.

The model we employ is the same as in \cite{SZL}. We model an mRNA with $N$
codons as a chain of sites, each of which is labeled by $i$. The first and last
sites, $i=1,N$, are associated with the start and stop codons, respectively. At
any time, attached to the mRNA are $M$ ribosomes (particles). Being a large
complex of molecules, each ribosome will cover $\ell $ sites (codons), with
$\ell =12$ typically \cite{Heinrich, Kang}. Any site may be covered by a single
ribosome or none. In case of the latter, we will refer to the site as ``empty''
or ``occupied by a hole.'' To locate the ribosome, we arbitrarily choose the
\emph{lowest} site covered. For example, if the first $\ell$ sites are empty, a
ribosome can bind in an initiation step, and then it is said to be ``on site
$i=1$.''  We define $n_i$ to be the ribosome density at site $i$, where only
the left end of the ribosome is counted.  (In \cite{MGP, MG}, particles were
located by their right end, but either choice leads to the same rules of
motion.)  We also define the coverage density $\rho_i=\sum_{s=0}^{\ell-1}
{n_{i-s}}$, which is the probability that site $i$ is covered by some portion
of a ribosome.

Next, we specify the dynamics of our model. An attached ribosome located at site $i$ will move to the next site ($i+1$) with a rate $k_{i}$,
provided site $i+\ell $ is empty. For Monte Carlo simulations, it is convenient to update configurations in discrete time units $\Delta t$. In
implementing the simulations, it is better to use probabilities $p_{i} = k_i \Delta t$, so that a ribosome on site $i$ will be moved (or not) with
probability $p_{i}$ (or $ 1-p_{i}$). We purposefully associate these hopping probabilities with a site because a site is associated with a
particular codon. Thus, the hopping rate from that site may depend on the relative abundance of the appropriate aa-tRNA. Apart from these
probabilities, another aspect of our simulations is random sequential updating: i.e., during each Monte Carlo step (MCS), $M+1$ particles are
chosen at random, in sequence, to attempt moves. They are selected from a pool that includes the $M$ particles on the lattice plus another unbound
particle that can initiate if there are $\ell $ holes at the beginning of the chain.

In our computational studies, systems begin empty and are run for long enough to reach steady state. After steady state is attained, data
including the current and density distribution can be collected.  Density data is typically collected every 100 MCS. We often use continuous time
Monte Carlo \cite{BKL} because it runs far more quickly than and provides the same results as standard Monte Carlo.  Using continuous time Monte
Carlo also avoids the need to specify a fixed time step $\Delta t$.

In our studies of real mRNA sequences, we use gene sequences from \textit{Escherichia coli} strain MG1655, obtained from \cite{MG1655}. Elongation
rates at each codon are estimated using commonly accepted values for the availability of tRNA in \textit{E. coli} \cite{Solom}. The rate at each
codon is assumed proportional (with an arbitrary proportionality constant) to the availability of the appropriate tRNA that can decode the codon,
as in \cite{Lesnik}. Corresponding data are not available for estimating initiation and termination rates, so we choose various rates of interest
to study.

\section{Summary of previous results}
\label{sec:literature}

Extensive investigations of the simple totally asymmetric exclusion process
(TASEP, defined as point particles hopping with unit rate along a line) with
open boundaries can be found in the literature.  We first consider studies of
uniform systems.  Exact analytic results for the $\ell=1$ steady state exist
\cite{DDM, SD}. Systems with extended objects ($\ell >1$) have been less
frequently investigated but have also been understood from various approaches.
Using a mean field approach, MacDonald \textit{et al.} derived mean field
equations for the average site occupation $ \left\langle n_i\right\rangle $ and
considered both closed \cite{MGP} and open \cite{MG} systems. In the former
case, exact solutions were found, leading to a current vs. density relation.
For the latter, the authors resorted to numerical solutions to find the phase
diagram for a variety of initiation and termination rates.  Lakatos and Chou
\cite{LC} considered uniform open systems with extended objects. Using a
discrete Tonks gas partition function, they derived the current vs. density
relation first presented by MacDonald \textit{et al.} \cite{MGP}. Via a refined
mean field theory, they extended this result to predict currents and bulk
densities for the open system, which they confirmed by Monte Carlo simulations.
Finally, Shaw \textit{et al.} \cite{SZL} used an extremal principle \cite{PS}
based on domain wall theory \cite{KSKS} to obtain the phase diagram, with
currents and bulk densities, for the uniform open system.  They also found
approximate density profiles (related to the coverage density $\rho$) from a
continuum limit.  From all of these studies, the $\ell>1$ phase diagram is well
known.  Depending on the initiation (or injection) and termination (or
depletion) rates, the system will settle into one of three phases. From their
dominant characteristics, the three phases are known as low density, high
density, and maximal current. The initiation and termination probabilities are
often referred to as $\alpha$ and $\beta$ in the literature. A phase diagram in
this $\alpha $-$\beta $ plane has been determined, showing second order
transitions between the maximal current phase and the others, as well as a
first order transition between the high- and low-density regions.

When disorder is introduced, i.e., not all the $p_i$'s are equal, then methods for exact analytic approaches fail (except in the extremely dilute
limit, where only the motion of a single particle is of concern \cite{Derrida}). Indeed, even a single slow rate in a \emph{closed} system poses
serious difficulties \cite{JL1, JL2, Sep}. However, Kolomeisky \cite{ABK} obtained approximate steady state solutions and phase diagrams for an
\emph{open} system with a single nonuniform rate in the bulk by splitting the system into two smaller systems connected by the nonuniform rate.
Tripathy and Barma \cite{Barma} considered a closed system, but with a finite fraction of identical slow sites. Based on a combination of Monte
Carlo simulations and numerical solutions of mean field equations, they found current-density relations. Further references on TASEP with disorder
may be found in a recent review \cite {JK-BJP}. All of these studies are restricted to $\ell =1 $.  Studies of disorder in systems with $\ell>1$
have been fairly limited.  Shaw \textit{et al.} \cite{SZL} found upper and lower bounds for the current in systems with arbitrary disorder. In
another work, Shaw \textit{et al.} \cite{SKL} considered an open system with $\ell>1$ and a single nonuniform rate in the bulk.  As was done for
$\ell=1$ \cite{ABK}, the system was divided into two smaller systems connected by the nonuniform rate.  Steady state currents and bulk densities
to either side of the nonuniform site were obtained.

\section{Simple coarse-graining approach}
\label{sec:coarsegrained}

We consider briefly an approximate method motivated by the method of Shaw
\textit{et al.} \cite{SKL}.  Their particle hopping rate in the bulk was $1$,
except for the nonuniform rate $q$ at special site $k$.  Important in their
analysis is the parameter
\begin{equation*}
q_{eff}=\frac{\ell}{1/q+\left(\ell-1\right)}.
\end{equation*}
This parameter appears in the current passing from the left sublattice into the right sublattice:
\begin{equation}
J=q_{eff} \frac{\rho_{left}}{\ell}
\frac{\left(1-\rho_{right}\right)}{\left(1-\rho_{right}+\rho_{right}/\ell\right)},
\label{Jatblockage}
\end{equation}
where $\rho_{left}$ and $\rho_{right}$ are the bulk densities in
the left and right sublattices.  We note that $q_{eff}$ is the
same as $\ell K_{\ell,k}$ in the notation of \cite{SZL}, where
\begin{equation*}
K_{\ell ,i}\equiv \left( \sum_{j=i}^{i+\ell -1}\frac 1{k_j}\right)
^{-1}
\end{equation*}
is a coarse-grained rate for translating $\ell$ sites.

\begin{figure}[tbp]
\includegraphics[clip=true]{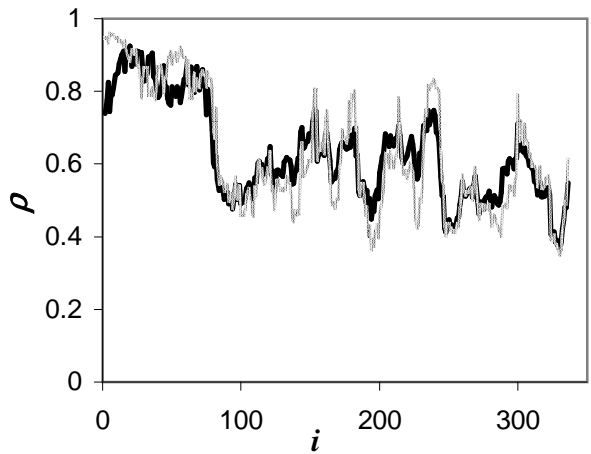}
\caption{Coverage density profile for the \textit{ompA} gene of \textit{E. coli} when elongation rates are limiting.  Bold curve is Monte Carlo
simulation result, and lighter curve is coarse-grained prediction from Eq. \ref{rhocoarsegrained} using the Monte Carlo current.  The positive
root was used to the left of the current-limiting region and the negative root to the right.  The real part is plotted where the predicted $\rho$
is complex. Elongation rates at each codon were assumed proportional to availabilities of corresponding tRNA \cite{Solom}.}
\label{fig:coarsegrained}
\end{figure}

The form of Eq. \ref{Jatblockage} motivates us to suggest that
\begin{equation}
J=\ell K_{\ell,i} \frac{\rho_i}{\ell}
\frac{1-\rho_i}{1-\rho_i+\rho_i/\ell}  \label{Jcoarsegrained}
\end{equation}
in regimes in which the coverage density $\rho$ is slowly varying in $i$.  Because $\rho$ and $K_{\ell}$ are both coarse-grained over a distance
$\ell$, a relationship between them is unsurprising. Eq. \ref{Jcoarsegrained} can be solved for $\rho_i$:
\begin{eqnarray}
\rho_i &=& \frac{1}{2K_{\ell,i}} \left[ K_{\ell,i} +J-J/\ell \vphantom{\pm \sqrt{ \left( K_{\ell,i}+J-J/\ell \right)^2 -4JK_{\ell,i}}} \right. \nonumber \\
 & & \left. \pm \sqrt{ \left( K_{\ell,i}+J-J/\ell \right)^2 -4JK_{\ell,i} } \right]
\label{rhocoarsegrained}
\end{eqnarray}
It is reasonable to use the positive (high density) root to the left of the
current-limiting region (where the minimum $K_{\ell}$ occurs) and the negative
(low density) root to the right. Results of this approach are shown in Fig.
\ref{fig:coarsegrained} for the \textit{ompA} gene when elongation rates are
limiting (i.e., initiation and termination rates are sufficiently large).  The
value for current $J$ in Eq. \ref{rhocoarsegrained} is taken from Monte Carlo
simulations of \textit{ompA}. The agreement between the coarse-grained result
and Monte Carlo simulations is good in low density regions but is poorer in
high density regions because long range correlations become more important, an
effect not captured by coarse-graining over the relatively short distance
$\ell$. Finally, we note that Eq. \ref{rhocoarsegrained} can be used only when
the current $J$ is known, either from Monte Carlo simulations or from some
analytical means.

\section{Mean field approach}
\label{sec:MGP}

We next turn to a mean field approach, using equations first developed by
MacDonald \textit{et al.} \cite{MGP, MG}.  The equations were applied only to
uniform systems, but we will find that they can be successfully applied to
nonuniform systems. Here the location of a particle is determined by the
location of its left end. The particle density at site $i$ is $n_i$, and the
hole density at site $i$ is $1-\sum_{s=0}^{\ell-1} {n_{i-s}}$. For a particle
to move from site $i$ to $i+1$, producing a current, site $i+\ell$ must be
empty. The method is ``mean field'' in the sense that some correlations have
been neglected. The conditional probability that site $i+\ell$ is empty given
that site $i$ contains a particle is replaced by the conditional probability
that site $i+\ell$ is empty given that site $i$ contains either a particle or a
hole. That is,
\begin{eqnarray*}
P\left(\bullet-\circ \right. &\mid& \left. \bullet-?\right) =
\frac{P(\bullet-\circ)}{P(\bullet-\circ) + P(\bullet-\bullet)} \nonumber \\
&\approx& P\left(?-\circ \mid ?-? \right) \nonumber \\ &=&
\frac{P(\bullet-\circ)+P(\circ-\circ)}{P(\bullet-\circ) +
P(\bullet-\bullet)+P(\circ-\circ)+P(\circ-\bullet)} \nonumber \\ &=&
\frac{1-\sum_{s=1}^{\ell}{n_{i+s}}}{1-\sum_{s=1}^{\ell}{n_{i+s}}+n_{i+\ell}},
\end{eqnarray*}
where we use $\bullet$ to denote a filled site and $\circ$ to denote an empty
site.

The mean field assumption for the conditional probability leads to the following equations for the time evolution of the $\left\{n_i\right\}$:
\begin{widetext}
\begin{eqnarray}
\frac{dn_1}{dt} &=& k_0 \left( 1-\sum_{s=1}^{\ell}{n_s} \right) - k_1 n_1 \frac{1-\sum_{s=1}^{\ell}{n_{1+s}}}{1-\sum_{s=1}^{\ell}{n_{1+s}}+n_{1+\ell}} \nonumber \\
\frac{dn_i}{dt} &=& k_{i-1} n_{i-1} \frac{1-\sum_{s=1}^{\ell}{n_{i-1+s}}}{1-\sum_{s=1}^{\ell}{n_{i-1+s}}+n_{i-1+\ell}} - k_i n_i
\frac{1-\sum_{s=1}^{\ell}{n_{i+s}}}{1-\sum_{s=1}^{\ell}{n_{i+s}}+n_{i+\ell}} \text{  for $i=2,\ldots,N-\ell$} \nonumber \\
\frac{dn_{N-\ell+1}}{dt} &=& k_{N-\ell} n_{N-\ell} \frac{1-\sum_{s=1}^{\ell}{n_{N-\ell+s}}}{1-\sum_{s=1}^{\ell}{n_{N-\ell+s}}+n_{N}} -
k_{N-\ell+1}
n_{N-\ell+1} \nonumber \\
\frac{dn_i}{dt} &=& k_{i-1} n_{i-1} - k_i n_i \text{  for $i=N-\ell+2,\ldots,N$}, \label{MFdiffeq}
\end{eqnarray}
\end{widetext}
where we use $k_0$ to denote the initiation rate.

For the steady state solution, time derivatives are set to $0$ and
the current $J$ is introduced.  The resulting set of equations,
\begin{subequations}
\label{MFiter}
\begin{eqnarray}
J &=& k_0 (1-\sum_{s=1}^{\ell}{n_s}) \label{IC} \\
n_i &=& \frac{J}{k_i}
\frac{1-\sum_{s=1}^{\ell}{n_{i+s}}+n_{i+\ell}}{1-\sum_{s=1}^{\ell}{n_{i+s}}} \nonumber \\
& & \quad \quad \quad \text{  for $i=1,\ldots,N-\ell$} \label{bulk} \\
n_i &=& \frac{J}{k_i} \text{  for $i=N-\ell+1,\ldots,N$}, \label{term}
\end{eqnarray}
\end{subequations}
can be solved iteratively for $n_N$ to $n_1$ if $J$ is specified. Then Eq.
\ref{IC} becomes an initial condition to check for consistency to determine
whether $J$ has been chosen correctly. If Eq. \ref{IC} is not satisfied, then
$J$ should be adjusted appropriately and the process repeated.

We present an argument for the validity of this iterative approach. First, a
physically meaningful solution will have particle density $n_i \in
\left(0,1\right)$ and coverage density $\sum_{s=1}^{\ell}{n_{i+s}} \in
\left(0,1\right)$ for all $i$. (Endpoints of the interval are excluded if there
is to be nonzero current flow.) Suppose that for some $i$, $J
> k_i (1-\sum_{s=1}^{\ell}{n_{i+s}})$. Then from Eq. \ref{bulk}, $n_i >
1-\sum_{s=1}^{\ell-1}{n_{i+s}}$, meaning that $\sum_{s=0}^{\ell-1}{n_{i+s}}
>1$, which is a contradiction.  So we see that $J \leq k_i
(1-\sum_{s=1}^{\ell}{n_{i+s}})$ for all $i$.  (Note that although we have not
dealt separately with $i=N-\ell+1,\ldots,N$, Eq. \ref{term} is consistent with
the previous statement.)  Thus, $J$ cannot be too large if physical solutions
are to be obtained.

Next, we show that the densities $\left\{n_i\right\}$, while within physical
ranges, are increasing functions of $J$. Consider two different $J$ values:
$J_0$ with its associated densities $\left\{n_i\right\}$ and $J_1$ with its
densities $\left\{m_i\right\}$, which we assume to be in physical ranges.
Suppose that $J_1>J_0$.  Clearly $m_i > n_i$ for $i=N-\ell+1,\ldots,N$. Using
induction on the remaining $i$, it can be shown from Eq. \ref{bulk} that
\begin{widetext}
\begin{eqnarray*}
m_i-n_i &>& \frac{J_0}{k_i}
\frac{1-\sum_{s=1}^{\ell}{m_{i+s}}+m_{i+\ell}}{1-\sum_{s=1}^{\ell}{m_{i+s}}} -
\frac{J_0}{k_i}
\frac{1-\sum_{s=1}^{\ell}{n_{i+s}}+n_{i+\ell}}{1-\sum_{s=1}^{\ell}{n_{i+s}}}
\nonumber \\
&=& \frac{J_0}{k_i \left( 1-\sum_{s=1}^{\ell}{m_{i+s}}\right)
\left(1-\sum_{s=1}^{\ell}{n_{i+s}} \right) } \left[ m_{i+\ell}
\left(1-\sum_{s=1}^{\ell}{n_{i+s}} \right) - n_{i+\ell} \left(
1-\sum_{s=1}^{\ell}{m_{i+s}}\right) \right] \nonumber \\
&>& 0. \nonumber
\end{eqnarray*}
\end{widetext}
Therefore, the densities $\left\{n_i\right\}$ increase with increasing $J$.

Finally, we again consider Eq. \ref{IC}.  The left side increases monotonically
with increasing $J$, and the right decreases monotonically with increasing $J$,
while densities are in physical ranges. If we follow the iterative approach,
$J$ values that lead to $\sum_{s=1}^{\ell}{n_{i+s}}>1$ or $J>k_0 \left(
1-\sum_{s=1}^{\ell}{n_s} \right)$ are too large and should be decreased.  On
the other hand, $J$ values that lead to $J<k_0 \left( 1-\sum_{s=1}^{\ell}{n_s}
\right)$ are too small and should be increased.  One can start with upper and
lower bounds for the current (from \cite{SZL}) and use bisection to converge to
the correct current.  If a physical solution exists, it is unique and should be
found by this method.

Note that the upper bound for $J$ from \cite{SZL},
\begin{equation}
J \leq \left( \sum_{s=0}^{\ell-1}{\frac{1}{k_{i+s}}}\right)
\end{equation}
for all $i$, applies also to the mean field equations.  This can
be shown from
\begin{eqnarray*}
\sum_{s=1}^{\ell}{\frac{J}{k_{i+s}}} &=& \sum_{s=1}^{\ell}{n_{i+s}}
\frac{1-\sum_{t=1}^{\ell}{n_{i+s+t}}}{1-\sum_{t=1}^{\ell}{n_{i+s+t}}+n_{i+s+\ell}}
\nonumber
\\
&\leq& \sum_{s=1}^{\ell}{n_{i+s}} \leq 1. \nonumber
\end{eqnarray*}

\begin{figure}[tbp]
\includegraphics[clip=true]{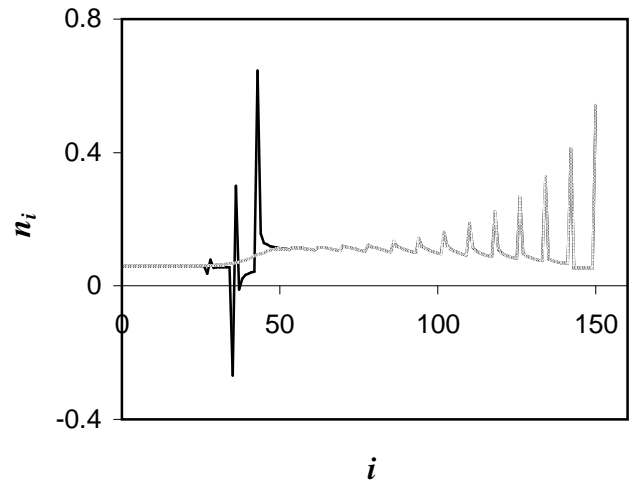}
\caption{Particle density profiles calculated by iteration of mean field
equations (Eq. \ref{MFiter}) for a uniform system with initiation rate 1,
elongation rates 1, and termination rate 0.1. The system had $N=150$ and
$\ell=8$. The dark curve is the result for current $J$ slightly too large, and
the light/diffuse curve is the result for $J$ slightly too small. The two
curves are superimposed for $i>50$. The difference between the two $J$ values
was $2 \times 10^{-19} J$. } \label{fig:MFdens}
\end{figure}

In practice, computing iterative solutions for $\left\{n_i\right\}$ and
adjusting $J$ by bisection is effective in finding $J$ and finding $n_i$ values
that are low density (downstream of the current-limiting region).  Table
\ref{tab:currents} shows that there is fairly good agreement (within 5\%)
between the mean field current and the actual current (from Monte Carlo
simulations) for various real gene sequences of \textit{E. coli}. However,
numerical problems arise in finding $n_i$ values that are high density
(upstream of the current-limiting region). For high density solutions, we have
observed that there generally exists a very narrow range for $J$, with a width
less than machine precision, for which the smaller $J$ values will fail to
satisfy Eq. \ref{IC} because the densities are too small, and for which the
larger $J$ values will lead to nonphysical densities. An example of this
phenomenon is shown in Fig. \ref{fig:MFdens}.

\begin{table}
\caption{\label{tab:currents}Currents for real gene sequences of \textit{E.
coli} from Monte Carlo (MC) simulations and mean field (MF) calculations. Both
the original mean field and the two-point mean field are included. Units for
the currents are arbitrary.  Percent errors for the mean field currents, as
compared with the Monte Carlo currents, are given in parentheses. Elongation
rates were assumed proportional to the availability of the appropriate tRNA for
each codon \cite{Solom}. Several of the genes were chosen to be
initiation-limited by making the initiation rate 0.78 of the slowest elongation
rate. Others were made termination-limited by making the termination rate 0.34
(for asnS and envY) or 0.52 (for fabG) of the slowest elongation rate. The
remainder were elongation-limited. Errors in the Monte Carlo results are less
than 0.001.}
\begin{ruledtabular}
\begin{tabular}{ccccc}
gene & limit & MC $J$ & orig. MF $J$ (\%err) & 2-pt. MF $J$ (\%err) \\
\hline

adk & elong. & 0.139 & 0.133 (4.3) & 0.137 (1.4) \\
cysK & elong. & 0.120 & 0.122 (1.7) & 0.118 (1.7) \\
gapA & elong. & 0.194 & 0.191 (1.5) & 0.191 (1.5) \\
glnH & elong. & 0.156 & 0.154 (1.3) & 0.158 (1.3) \\
aceF & init. & 0.170 & 0.164 (3.5) & 0.166 (2.4) \\
crr & init. & 0.172 & 0.177 (2.9) & 0.172 (0.0) \\
fabD & init. & 0.114 & 0.112 (1.8) & 0.112 (1.8) \\
asnS & term. & 0.114 & 0.114 (0.0) & 0.114 (0.0) \\
envY & term. & 0.092 & 0.091 (1.1) & 0.091 (1.1) \\
fabG & term. & 0.112 & 0.113 (0.9) & 0.112 (0.0)
\end{tabular}
\end{ruledtabular}
\end{table}

We next present an argument for why high density solutions are associated with
numerical difficulties.  For convenience, we assume that the $k_i$ are
uniformly $1$, and we seek uniform density solutions.  Eq. \ref{bulk} gives an
iterative map for $n_i$. We assume that a fixed point $n^{\ast}$ exists. It
will then satisfy
\begin{equation*}
n^{\ast }=J\frac{1-\left( \ell-1\right) n^{\ast }}{1-\ell n^{\ast }}.
\end{equation*}
We find high density and low density fixed points,
\begin{equation*}
n_{\pm }^{\ast }=\frac{1}{2\ell}\left[ 1+J\left( \ell-1\right) \pm
\sqrt{\left[ 1+J\left( \ell-1\right) \right] ^{2}-4\ell J}\right].
\end{equation*}

Suppose the densities $n_{i+1},\ldots,n_{i+\ell}$ are slightly perturbed from the high density fixed point, so that $n_{j}=n_{+}+\delta $, where
$\left| \delta \right| \ll 1$, for $j=i+1,...,i+\ell $. Then
\begin{eqnarray*}
n_{i} &=&J\frac{1-\left( \ell-1\right) n_{+}-\left( \ell-1\right)
\delta }{
1-\ell n_{+}-\ell \delta } \\
&=&n_{+}+\frac{4J}{\left[ -1+\left( \ell-1\right) J+\sqrt{\left[ 1+J\left(
\ell-1\right) \right] ^{2}-4\ell J}\right] ^{2}}\delta \nonumber \\
&& \quad \quad +O\left( \delta ^{2}\right) .
\end{eqnarray*}

To determine stability of the high density fixed point, we
consider the $\delta $ prefactor:
\begin{equation*}
a\equiv \frac{4J}{\left[ -1+\left( \ell-1\right) J+\sqrt{\left[
1+J\left( \ell-1\right) \right] ^{2}-4\ell J}\right] ^{2}}
\end{equation*}
For currents in the expected range, $J \in \left(0,1/(1+\sqrt{\ell})^2\right)$
(cf. \cite{MG}), $a>1$ so that the high density fixed point is unstable. A
similar argument shows the stability of the low density fixed point.  Although
these ideas are proven here only for uniform density solutions, our numerical
results (such as those in Fig. \ref{fig:MFdens}) lead us to believe that the
nonuniform density case is similar. It appears that small numerical
imprecisions prevent us from accurately finding high density solutions, while
low density solutions can be easily found.

\begin{figure}[tbp]
\includegraphics[clip=true]{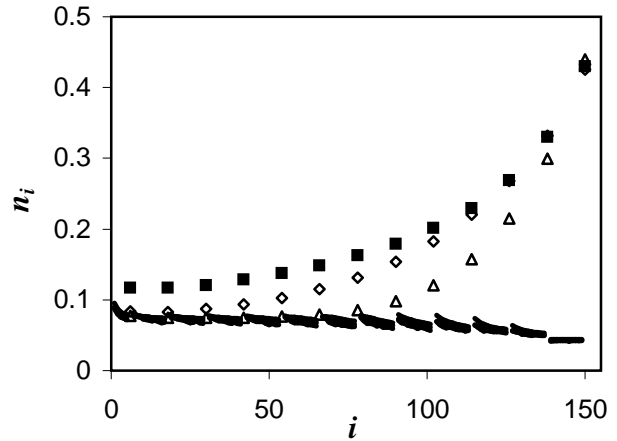}
\caption{Particle density profile for a uniform system with initiation rate 1,
elongation rates 1, and termination rate 0.1.  The system had $N=150$ and
$\ell=12$.  Density peaks (every $\ell$ sites) are displayed as symbols: filled
squares are the Monte Carlo simulation result, open triangles are the
prediction from the original mean field theory, and open diamonds are the
prediction from the two-point mean field theory. Non-peak densities are
displayed as curves; non-peak results from all three methods overlap.}
\label{fig:HDprofile}
\end{figure}

Finding steady state high density mean field solutions is a nontrivial problem.
We have attempted multidimensional Newton's method approaches to solve Eq.
\ref{MFiter} for $\left\{n_i\right\}$ and $J$, but these have their own
difficulties. Convergence often fails unless the initial guess is very near the
solution.  The most reliable method is to start with an empty lattice and
integrate Eq. \ref{MFdiffeq} to steady state. Although integration is
computationally slow, it consistently produces density profiles that are
reasonable and similar to the Monte Carlo simulation density profiles.
Agreement is best in the low density regime, when the correlations neglected by
the mean field theory are less important (data not shown). In the high density
regime, the mean field results underestimate the correlations between adjacent
particles. An example of this discrepancy is shown for a uniform system in Fig.
\ref{fig:HDprofile}.

\section{Mean field with two-point correlations}
\label{sec:twopoint}

To obtain more accurate solutions for density profiles, especially in the high
density regime, we extend the mean field theory to include two-point
correlations between adjacent sites. Variables in the two-point mean field
theory are densities of bonds, indexed by $i$, where bond $i$ connects site $i$
to site $i+1$.  There are four types of bond densities, which we denote as
follows: $n_{\circ,i}$, hole-hole pairs ($\circ-\circ$); $n_{\rightarrow,i}$,
particle-hole pairs ($\bullet-\circ$); $n_{\leftarrow,i}$, hole-particle pairs
($\circ-\bullet$); and $n_{\times,i}$, particle-particle pairs
($\bullet-\bullet$).  Geometry requires that
\begin{equation}
n_{\times,i}+n_{\leftarrow,i}=n_{\times,i+\ell}+n_{\rightarrow,i+\ell}
\label{geomparticle}
\end{equation}
and
\begin{equation}
n_{\circ,i}+n_{\rightarrow,i}=n_{\circ,i+1}+n_{\leftarrow,i+1}.
\label{geomhole}
\end{equation}
A third equation,
\begin{equation*}
1=n_{\rightarrow,i}+n_{\circ,i}+\sum_{s=1}^{\ell}{\left(n_{\times,i+s}+n_{\leftarrow,i+s}\right)},
\end{equation*}
can be written because each lattice site is occupied by either a
hole or some part of a particle.  However, this third equation is
linearly dependent on Eqs. \ref{geomparticle} and \ref{geomhole}.

We can use Eqs. \ref{geomparticle} and \ref{geomhole} to eliminate $n_{\leftarrow,i}$ and $n_{\circ,i}$ from the problem, so we will write
differential equations for the time evolution of only the two remaining types of densities, $n_{\rightarrow,i}$ and $n_{\times,i}$.  Fluxes into
and out of each state take the form of a product of the appropriate rate constant, the density of the state that may evolve, and the conditional
probability that adjacent particles and holes are arranged appropriately for the evolution to occur.  For example, a hole-hole pair at bond $i$
will evolve to a particle-hole pair at bond $i$ with rate
\begin{equation*}
k_{i-1} n_{\circ,i} P(\bullet-\circ-\circ \mid ?-\circ-\circ).
\end{equation*}
We make mean field assumptions for the conditional probability, similar to that of MacDonald \textit{et al.} \cite{MGP}.  For example,
\begin{eqnarray*}
P(\bullet-\circ-\circ &\mid& ?-\circ-\circ) \nonumber \\
&\approx& P(\bullet-\circ-? \mid ?-\circ-?) \nonumber \\
&=& \frac{n_{\rightarrow,i-1}}{n_{\rightarrow,i-1}+n_{\circ,i-1}}.
\end{eqnarray*}
Thus the flux of hole-hole pairs at site $i$ to particle-hole pairs at site $i$ is
\begin{equation*}
k_{i-1} n_{\circ,i} \frac{n_{\rightarrow,i-1}}{n_{\rightarrow,i-1}+n_{\circ,i-1}}
\end{equation*}

The resulting differential equations for time evolution of the bond densities
in the bulk are
\begin{widetext}
\begin{subequations}
\label{2ptdiffeq}
\begin{eqnarray}
\frac{dn_{\rightarrow,i}}{dt} &=& k_{i+\ell} n_{\times,i} \frac{n_{\rightarrow,i+\ell}}{n_{\rightarrow,i+\ell}+n_{\times,i+\ell}} + k_{i-1}
n_{\circ,i} \frac{n_{\rightarrow,i-1}}{n_{\rightarrow,i-1}+n_{\circ,i-1}} -k_i n_{\rightarrow,i} \label{nr} \\
\frac{dn_{\times,i}}{dt} &=& k_{i-1} n_{\leftarrow,i}
\frac{n_{\rightarrow,i-1}}{n_{\rightarrow,i-1}+n_{\circ,i-1}} - k_{i+\ell}
n_{\times,i}
\frac{n_{\rightarrow,i+\ell}}{n_{\rightarrow,i+\ell}+n_{\times,i+\ell}}.
\label{nx}
\end{eqnarray}
\end{subequations}
We also have boundary conditions, and for convenience, we assume that particles enter the lattice one site at a time, so that only the first site
must be free for initiation to occur.  We also assume that a particle whose right edge is on site $N$ can leave the lattice, freeing the final
$\ell$ sites.  Particle-particle bonds thus cannot exist in the final $\ell$ sites.  Then the boundary conditions are, for initiation,
\begin{eqnarray*}
\frac{dn_{\rightarrow,1}}{dt} &=& k_{1+\ell} n_{\times,1} \frac{n_{\rightarrow,1+\ell}}{n_{\rightarrow,1+\ell}+n_{\times,1+\ell}} + k_{0}
n_{\circ,1} -k_1 n_{\rightarrow,1} \\
\frac{dn_{\times,1}}{dt} &=& k_{0} n_{\leftarrow,1} - k_{1+\ell} n_{\times,1}
\frac{n_{\rightarrow,1+\ell}}{n_{\rightarrow,1+\ell}+n_{\times,1+\ell}}
\end{eqnarray*}
and, for $i=N-\ell+1,\ldots,N$,
\begin{eqnarray*}
\frac{dn_{\rightarrow,i}}{dt} &=& k_{i-1} n_{\circ,i} \frac{n_{\rightarrow,i-1}}{n_{\rightarrow,i-1}+n_{\circ,i-1}} -k_i n_{\rightarrow,i} \\
\frac{dn_{\times,i}}{dt} &=& 0.
\end{eqnarray*}
\end{widetext}

It is possible to obtain an iterative steady state solution for the bond
densities, from $i=N$ to $i=1$.  However, this solution appears to exhibit
numerical instabilities in the high density regime that are similar to those
observed with the original mean field theory.  We would prefer to have a simple
method, like the iteration and bisection method, for computing the current
despite the numerical difficulties in computing the densities. However, such a
method is not apparent.  Instead, we begin with an empty lattice and directly
integrate the differential equations for the bond densities (Eq.
\ref{2ptdiffeq}) until steady state is attained.  The two-point mean field
approach produces both densities and currents that agree more closely with
Monte Carlo simulations than did the original (one-point) mean field theory.
Table \ref{tab:currents} compares two-point mean field currents with currents
from Monte Carlo and the one-point mean field theory for real gene sequences.
In each case, the two-point mean field does as well as or better than the
original mean field at matching the Monte Carlo currents. Fig.
\ref{fig:HDprofile} compares the two-point mean field density profile with that
obtained by the other methods, showing that the two-point mean field theory
successfully incorporates more of the long-range correlations than does the
one-point mean field theory.

\section{Conclusions}
\label{sec:conclusions}

We have considered one-dimensional driven lattice gas models with large particles and quenched disorder.  Mean field theories were found be
effective in computing quantities of interest.  The mean field equations originally proposed by MacDonald \textit{et al.} \cite{MGP,MG} for
uniform systems were found to work equally well for nonuniform systems.  An iterative method allowed easy and rapid computation of the steady
state current through the system.  Steady state particle densities were also computed by this method, although only when the system was in a
low-density phase. We have gained some insight into the numerical difficulties that arise in obtaining high density solutions.  Direct integration
of differential equations for the time evolution of particle densities can always be used to find the steady state densities.  We found good
agreement between the mean field current and the Monte Carlo current.  Agreement between the densities was also adequate, though not as good in
the high density regime.

We extended the mean field approach to two-point correlations, using similar mean field approximations for conditional probabilities.  Currents
and particle densities were obtained more accurately from the two-point mean field theory than from the original.

Although the two-point mean field theory is an improvement on the original
theory, Fig. \ref{fig:HDprofile} shows that it still does not capture all of
the correlations necessary to reproduce high density profiles.  The theory
could be further extended to include three-point correlations (such as the
density of particle-particle-hole triplets).  However, the number of
independent variables and equations, as well as the complexity of the
equations, would increase as more correlations are added.  Also, numerical
instabilities might still exist so that solutions could be found only by
integration.  We therefore feel that it is not convenient to extend the method
to include higher-order correlations.

We conclude that mean field approaches can be effective in studying disordered
systems.  If one is primarily interested in the current through the system,
this quantity can be computed rapidly using iteration and bisection. We expect
the iteration and bisection method to be quite valuable in future studies,
because the calculated protein production rates could be compared to
experimental data (e.g., data in \cite{B&B}) and used in fitting of unknown
rate constants.

\begin{acknowledgments}
We thank Royce Zia for his helpful suggestions. KHL acknowledges support from
the NSF through grants BES-0120315 and 9874938. LBS was supported by a Corning
Foundation Fellowship and an IBM Ph.D. Fellowship.
\end{acknowledgments}

\end{document}